\documentclass[aps,prc,twocolumn,showpacs,tightenlines,
superscriptaddress]{revtex4}
\usepackage{epsfig,rotating}
\usepackage{multirow}

\newcommand{\nuc}[2]{\hbox{$^{#1}$#2}}

\begin{document}

\title{Spectroscopy of the odd-odd $fp$-shell nucleus \nuc{52}{Sc}
from secondary fragmentation}

\author{A.\ Gade}
   \affiliation{National Superconducting Cyclotron Laboratory,
      Michigan State University, East Lansing, Michigan 48824}
\author{R.\,V.\,F.\ Janssens}
    \affiliation{Physics Division, Argonne National Laboratory, Argonne,
      IL 60439}
\author{D.\ Bazin}
    \affiliation{National Superconducting Cyclotron Laboratory,
      Michigan State University, East Lansing, Michigan 48824}
\author{B.\,A.\ Brown}
    \affiliation{National Superconducting Cyclotron Laboratory,
      Michigan State University, East Lansing, Michigan 48824}
    \affiliation{Department of Physics and Astronomy,
      Michigan State University, East Lansing, Michigan 48824}
\author{C.\,M.~Campbell}
    \affiliation{National Superconducting Cyclotron Laboratory,
      Michigan State University,
      East Lansing, Michigan 48824}
    \affiliation{Department of Physics and Astronomy,
      Michigan State University, East Lansing, Michigan 48824}
\author{M.\,P.\ Carpenter}
    \affiliation{Physics Division, Argonne National Laboratory, Argonne,
      IL 60439}
\author{J.\,M.\ Cook}
    \affiliation{National Superconducting Cyclotron Laboratory,
      Michigan State University, East Lansing, Michigan 48824}
    \affiliation{Department of Physics and Astronomy,
      Michigan State University, East Lansing, Michigan 48824}
\author{A.\,N. Deacon}
    \affiliation{Department of Physics and Astronomy, Schuster Laboratory,
      University of Manchester, Manchester M13 9PL, United Kingdom}
\author{D.-C.\ Dinca}
    \affiliation{National Superconducting Cyclotron Laboratory,
      Michigan State University, East Lansing, Michigan 48824}
    \affiliation{Department of Physics and Astronomy,
      Michigan State University, East Lansing, Michigan 48824}
\author{S.\,J.\ Freeman}
    \affiliation{Department of Physics and Astronomy, Schuster Laboratory,
      University of Manchester, Manchester M13 9PL, United Kingdom}
\author{T.\ Glasmacher}
    \affiliation{National Superconducting Cyclotron Laboratory,
      Michigan State University, East Lansing, Michigan 48824}
    \affiliation{Department of Physics and Astronomy,
      Michigan State University, East Lansing, Michigan 48824}
\author{B.\,P.\ Kay}
    \affiliation{Department of Physics and Astronomy, Schuster Laboratory,
      University of Manchester, Manchester M13 9PL, United Kingdom}
\author{P.\,F.\ Mantica}
    \affiliation{National Superconducting Cyclotron Laboratory,
      Michigan State University,
      East Lansing, Michigan 48824}
    \affiliation{Department of Chemistry, Michigan State University,
      East Lansing, MI 48824}
\author{W.\,F.\ Mueller}
    \affiliation{National Superconducting Cyclotron Laboratory,
      Michigan State University, East Lansing, Michigan 48824}
\author{J.\,R.\ Terry}
    \affiliation{National Superconducting Cyclotron Laboratory,
      Michigan State University,
      East Lansing, Michigan 48824}
    \affiliation{Department of Physics and Astronomy,
      Michigan State University, East Lansing, Michigan 48824}
\author{S.\ Zhu}
    \affiliation{Physics Division, Argonne National Laboratory, Argonne,
      IL 60439}

\date{\today}

\begin{abstract}
The odd-odd $fp$-shell nucleus \nuc{52}{Sc} was investigated using
in-beam $\gamma$-ray spectroscopy following secondary fragmentation of
a \nuc{55}{V} and \nuc{57}{Cr} cocktail beam. Aside from the known
$\gamma$-ray transition at 674(5)~keV, a new decay at
$E_{\gamma}=212(3)$~keV was observed. It is attributed to the
depopulation of a low-lying excited level. This new state is discussed
in the framework of shell-model calculations with the GXPF1, GXPF1A,
and KB3G effective interactions. These calculations are found to be
fairly robust for the low-lying level scheme of \nuc{52}{Sc}
irrespective of the choice of the effective interaction. In addition,
the frequency of spin values predicted by the shell model is
successfully modeled by a spin distribution formulated in a
statistical approach with an empirical, energy-independent spin-cutoff
parameter.
\end{abstract}

\pacs{23.20.Lv, 21.60.Cs, 25.70.Mn, 27.40.+z} \keywords{\nuc{52}{Sc},
level scheme} \maketitle In-beam $\gamma$-ray spectroscopy with
intermediate-energy exotic beams provides a versatile tool to study
various aspects of nuclear structure beyond the valley of $\beta$
stability. While intermediate-energy Coulomb excitation of the
projectile is used to assess the degree of collectivity within an
exotic nuclear system~\cite{Gla98}, direct reactions -- one- and
two-nucleon knockout from the projectile \cite{Han03,Tos04} -- are
exploited to probe single-particle degrees of freedom. The evolution
and occupation of specific orbits within the nucleus can be tracked
with this method~\cite{Bro02}. Secondary fragmentation, in which
multiple nucleons are removed from the projectile, but not necessarily
in a direct reaction process, lacks the selectivity alluded to above
and, as a result, provides access to a wider variety of excited
states~\cite{Soh02}.

Nuclear structure of exotic species has been found to depart often
from expectations based on the properties of nuclei closer to
stability. New shell gaps appear~\cite{Oza00,Hon02,Jan02,Din05} and
"traditional" magic numbers vanish in the regime of pronounced
asymmetry between proton and neutron numbers (e.g.
\cite{WBB,Mot95,Gla97,Cot98,Nav00,Uts04,Bro05}). Those changes are
driven, for example, by the tensor force~\cite{Ots05} and by the
proton-neutron monopole interaction (for a recent reference on this
topic, see, e.g.,~\cite{Ots01}). The predictive power
of nuclear structure models is at present quite limited for exotic
nuclei, and shell-model interactions are adjusted by exploiting new
experimental data as they become available. For example, the GXPF1
effective interaction~\cite{Hon02}, which was optimized for nuclei in
the $fp$ shell, was modified recently following comparisons with new
experimental observations~\cite{Lid04,For04} in neutron-rich nuclei
just above $^{48}$Ca, which pointed to the need to adjust matrix
elements involving the $p_{1/2}$ orbital. The modified interaction has
been labeled GXPF1A~\cite{Hon04}.

We report here on the first observation of a low-energy $\gamma$ ray
in \nuc{52}{Sc} following secondary fragmentation of \nuc{55}{V} and
\nuc{57}{Cr}.  This transition had not been observed before and
presumably depopulates a low-lying excited state in this $fp$-shell
nucleus. Our experimental result is compared to shell-model
calculations using three effective interactions (GXPF1, GXPF1A and
KB3G) suited for the $fp$ shell. Furthermore, the shell-model
calculations with the GXPF1 effective interaction were probed further
by analyzing the frequency of spin values below the neutron separation
energy (73 states with $E_x \leq S_n=5.23$~MeV). It is shown that this
frequency can be described satisfactorily within a parameter-free,
statistical approach using an empirical, energy-independent
spin-cutoff parameter.

Previous knowledge about excited states of this nucleus stems from the
$\beta$ decay of the ground state of \nuc{52}{Ca}~\cite{Huc85}.
Consistent with the selection rules of this decay mode, only excited
states with $J^{\pi}$ assignments 1$^+$,($2^+$) have been
reported~\cite{Huc85}. The ground state is proposed to have tentative
spin and parity quantum numbers of $3^+$ based on the population
pattern of \nuc{52}{Ti} excited levels in the $\beta$ decay of the
ground state of \nuc{52}{Sc}~\cite{Huc85}.

The secondary beam cocktail was produced by fast fragmentation of a
130~MeV/nucleon \nuc{76}{Ge} primary beam delivered by the Coupled
Cyclotron Facility at the National Superconducting Cyclotron
Laboratory on a \nuc{9}{Be} primary target of
423~mg/cm$^2$ thickness. The
fragmentation products were selected in the A1900 fragment
separator~\cite{a1900}, which was operated at full momentum
acceptance. The cocktail beam containing \nuc{55}{V} and \nuc{57}{Cr}
with an average mid-target energy of 77 MeV/nucleon interacted with a
375~mg/cm$^2$ \nuc{9}{Be} foil placed in the target position of
the large-acceptance S800 spectrograph ~\cite{s800}. The reaction
residues were identified on an event-by-event basis from the
energy-loss measured in the S800 ionization chamber, the
time-of-flight measured between plastic scintillators, and the
position and angle information obtained with the two
position-sensitive cathode-readout drift chambers of the S800 focal
plane~\cite{s800}. \nuc{52}{Sc} residues produced from either the
fragmentation of \nuc{55}{V} or \nuc{57}{Cr} could not be disentangled
since the reaction products for those two constituents of the cocktail
beam overlapped in time of flight. The magnetic field of the
spectrograph was set to center two-proton knockout residues in the
focal plane, as these were the main focus of the
measurements~\cite{Ga06}. The large acceptance of the device allowed a
fraction of the \nuc{9}{Be}(\nuc{55}{V},\nuc{52}{Sc})X and
\nuc{9}{Be}(\nuc{57}{Cr},\nuc{52}{Sc})X residues to enter the S800
focal plane at the edge of the acceptance.

\begin{figure}[h]
        \epsfxsize 8.4cm
        \epsfbox{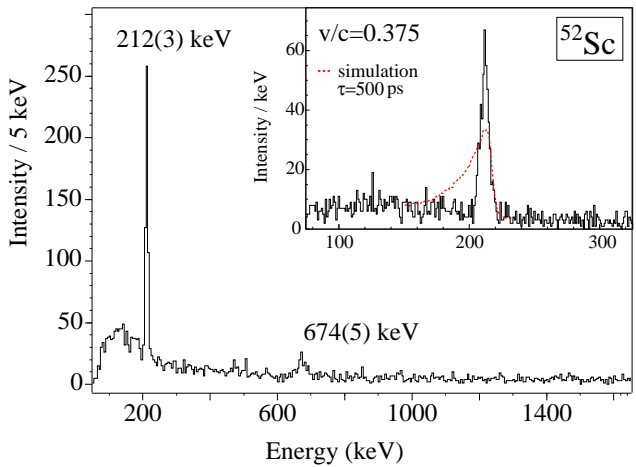}
\caption{\label{fig:spectrum} (Color online) Doppler-reconstructed
spectrum detected in coincidence with \nuc{52}Sc; the 674(5)-keV
transition was observed earlier~\cite{Huc85} while the prominent
212(3)-keV $\gamma$ ray is new. It originates from a ($4^+,5^+$) state
with a mean lifetime much shorter than 500~ps, as demonstrated in the
inset by a comparison with a simulation (dashed line). See text for
details.}
\end{figure}

The \nuc{9}{Be} reaction target was surrounded by SeGA, an array of
32-fold segmented HPGe detectors~\cite{sega}, arranged in two rings
with 90$^\circ$ and 37$^\circ$ central angles with respect to the beam
axis. The 37$^\circ$ ring was equipped with seven detectors while ten
detectors occupied the 90$^\circ$ positions. The $\gamma$ rays emitted by
fast-moving nuclei are detected with Doppler shifts in the laboratory
system. The high degree of segmentation of the SeGA detectors allows
for an event-by-event Doppler reconstruction where the angle of the
$\gamma$-ray emission is deduced from the position of the detector
segment that registered the highest energy deposition. The
event-by-event Doppler-reconstructed $\gamma$-ray spectrum in
coincidence with \nuc{52}{Sc} residues is shown in
Fig.~\ref{fig:spectrum}. A previously known $\gamma$-ray transition is
detected at 674(5)~keV. This transition was proposed in Ref.
~\cite{Huc85} to connect an excited $(2^+)$ state with the
\nuc{52}{Sc} ground state. The dominant peak in the spectrum, however,
corresponds to a $\gamma$-ray transition of 212(3)~keV, observed here
for the first time.

\begin{figure}[h]
        \epsfxsize 8.4cm
        \epsfbox{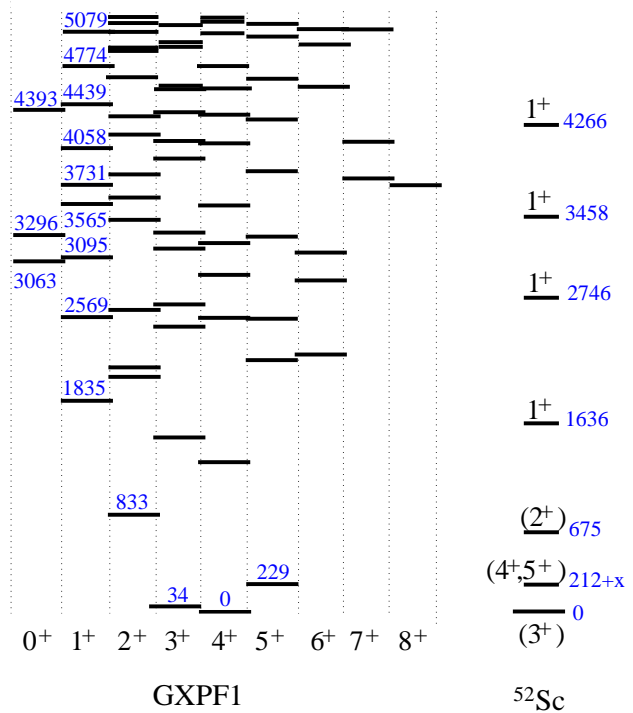}
\caption{\label{fig:level} (Color online) The results of the $fp$
shell-model calculations
  performed with the code
  {\sc oxbash} using the GXPF1~\cite{Hon02} effective interaction are
  compared with the
  experimental level scheme of \nuc{52}{Sc}. The
  calculations are shown up to the neutron separation energy. The
  experimental level scheme is taken from~\cite{Huc85}; a new level
  depopulated by the 212(3)-keV
  $\gamma$-ray transition reported here for the first time is
  discussed in the text in more detail.  }
\end{figure}

In Fig.~\ref{fig:level} the experimental level scheme known so far is
compared to the results of full shell-model calculations in the $fp$
model space employing the GXPF1 effective interaction. The {\sc
oxbash}~\cite{oxb85} calculations allowed for the 12 valence particles
with respect to the $^{40}$Ca core to occupy the $(f_{7/2}, p_{3/2},
f_{5/2}, p_{1/2})$ configuration space. The previously established
$(2^+)$ and $1^+$ levels are in good agreement with the calculations:
there is a one-to-one correspondence for the first excited $(2^+)$ and
for the first two $1^+$ states, while the high level density within
the shell model prevents a detailed comparison for levels above 3~MeV.
The main components of the shell-model wave functions obtained with
the GXPF1 interaction are given in Table~\ref{tab:sm_conf} for the
first $3^+$, $4^+$ and $5^+$ states. Configurations with the valence
protons and neutrons occupying the $f_{7/2}$ and $p_{3/2}$ orbitals
clearly dominate the structure of the low-lying states.

\begin{figure}[h]
        \epsfxsize 8.4cm
        \epsfbox{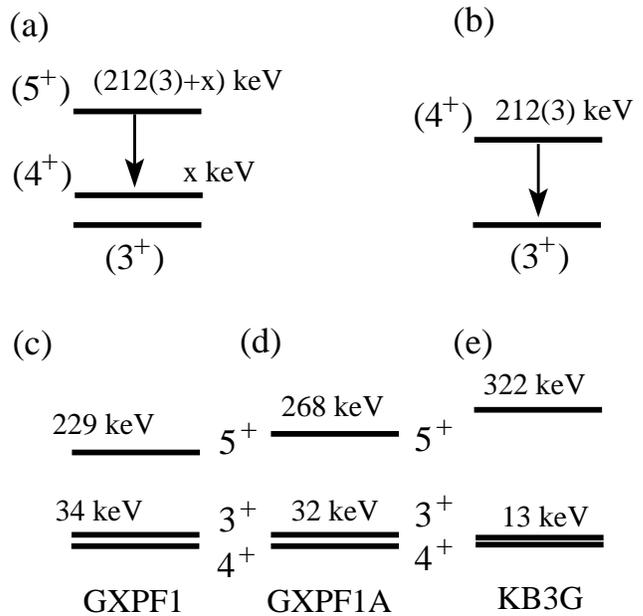}
\caption{\label{fig:cases} Two possible scenarios (a) and (b) for the
  placement of the 212(3)-keV transition in the decay scheme of
  \nuc{52}{Sc} compared with shell model calculations (c)--(e). An $E2$ transition from the $5^+$ to the $3^+$ level can be
  excluded as discussed in the text.}
\end{figure}

In the shell model, the first $4^+$ and $3^+$ states are almost
degenerate with the $4^+$ state becoming the ground state. Two
possible scenarios presented in Fig.~\ref{fig:cases} arise for the
placement of the newly observed low-energy $\gamma$-ray transition
within the level scheme of \nuc{52}{Sc}.  Considering the $(3^+)$
ground state suggested from $\beta$-decay studies~\cite{Huc85} and
guided by the shell-model calculations, the 212-keV $\gamma$ ray
either depopulates the first $5^+$ (Fig. 3(a)) or $4^+$ state (Fig.
3(b)). A $5^+ \rightarrow 3^+$ $E2$ transition can be excluded on the
basis of lifetime considerations. Indeed, only a lifetime of
$\tau>$550~ps would allow for the corresponding $B(E2;
5^+\rightarrow3^+)$ transition strength to be below the recommended
upper limit (RUL) of 300~W.u. for this mass region~\cite{End79}. Since
the Doppler reconstruction is very sensitive to the position of the
nucleus during $\gamma$-ray emission (due to its angle dependence), an
excited state with $\tau=550$~ps would decay roughly 5~cm behind the
reaction target. This would result in a pronounced low-energy tail for
the reconstructed photopeak. As shown in the inset of
Fig.~\ref{fig:spectrum}, the 212-keV line does not exhibit such an
asymmetry and, thus, an $E2$ transition can be excluded based on the
RUL. However, if the $5^+$ state decays to the $4^+$ level, presumably
almost degenerate with the ground state, the transition could proceed
with $M1$ character. Lifetimes of $\tau>1.1$~ps would conform with the
RUL for $M1$ transitions in this mass region. In the event of a near
degeneracy of the $4^+$ level with the ground state, the $4^+
\rightarrow 3^+$ decay would escape observation due to the detection
threshold (Fig.~\ref{fig:cases}(a)). The 212-keV transition could
possibly also connect the first excited $4^+$ to the $(3^+)$ ground
state, leading to the conclusion that the degree of degeneracy of the
first $4^+$ and $3^+$ states is overestimated in the shell-model
calculations (Fig.~\ref{fig:cases}(b)).

\begin{table}[h]
\begin{center}
 \vspace{0.5cm}
\caption{Dominant wave-function components $(f_{7/2})^{n_7}~(p_{3/2})^{n_3}~(f_{5/2})^{n_5}~(p_{1/2})^{n_1}$ for the first $3^+$, $4^+$, and $5^+$ states of \nuc{52}{Sc} from shell-model calculations with the GXPF1 effective interaction. Components with a strength below 1\% are not shown. }
\begin{ruledtabular}
\begin{tabular}{cccc}
$(n_7,n_3,n_5,n_1)$ & 3$^+_1$ & 4$^+_1$ & 5$^+_1$\\
                  & (\%) & (\%) & (\%) \\
\hline
(9,3,0,0)            &67.0 & 76.2&72.2 \\
(9,2,0,1)            &17.4 & 6.7 & 5.0 \\
(9,1,0,2)            &2.4 & 2.7  & 7.4  \\
(9,2,1,0)            &1.0 & 1.8 &   -   \\
(9,1,2,0)            & 1.2 & 1.3 & 2.1 \label{tab:sm_conf}
\end{tabular}
\end{ruledtabular}
\end{center}
\end{table}

Full $fp$ shell-model calculations have also been performed with the
GXPF1A and KB3G~\cite{Pov01} effective interactions in addition to
GXPF1.  The predicted low-lying level schemes can be found in
Fig.~\ref{fig:cases}(c,d,e). The results are very robust and support
experimental scenario (a). On the other hand, earlier calculations in
the full $fp$ shell with the FPD6 and KB3 effective interactions were
shown to differ significantly even for the low-lying
states~\cite{nov98}. However, the latter two interactions are known to
have shortcomings for neutron-rich nuclei in the region~\cite{For04}.
A truncated shell-model approach employing the TBLC8
interaction~\cite{Ric95} is closer to the calculations presented here.
The predictive power of the more recent and improved effective
interactions is demonstrated by the robustness of the calculations for
\nuc{52}{Sc}. Odd-odd nuclei are generally assessed to be very
sensitive to slight changes in the interaction~\cite{nov98}.

The shell-model calculations show, as expected, a rather high level
density for the odd-odd nucleus \nuc{52}{Sc}: a total of 73 states are
predicted below the neutron separation energy of $S_n=5.23$~MeV. The
number of states together with the dominance of the single-particle
degree of freedom make a Fermi-gas model applicable to the description
of the ``bulk'' properties of the excitation spectrum within the shell
model. These considerations prompted the study of the theoretical
level scheme with respect to statistical properties given below.

\begin{figure}[h]
        \epsfxsize 8.4cm
        \epsfbox{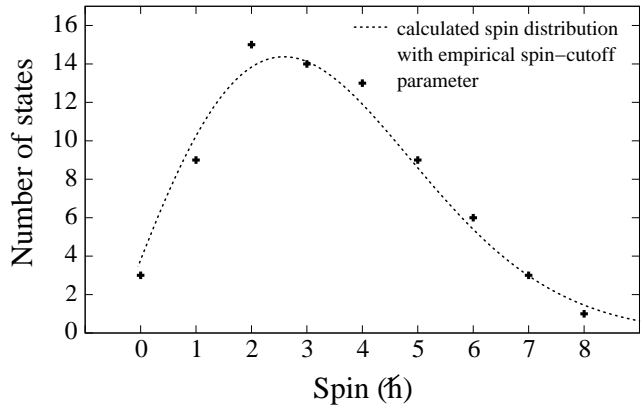}
\caption{\label{fig:density} Number of states per spin value in the
  shell-model calculations compared to the spin distribution modeled
  within the framework of
  a constant-temperature model with an empirical spin-cutoff
  parameter. The calculated spin distribution is drawn continuously
  to guide the eye. All 73 shell-model states with $E \leq
  S_n=5.23$~MeV are included.}
\end{figure}

Similar to the analyses presented in~\cite{Gar01,Egi88,Ign93,Gad02}
for experimental level schemes of heavier nuclei
closer to stability, a statistical
approach with an empirical spin-cutoff parameter was chosen to
describe the distribution of
spin values in the excitation spectrum of \nuc{52}{Sc} predicted within
the shell model. A separable
expression for the level density
$\rho(E,J)=\frac{1}{2}\rho(E)f(J)$ with an energy-dependent term $\rho(E)$
and the spin distribution $f(J)$,
\begin{equation}
f(J)\approx\frac{2J+1}{2\sigma^2}e^{-(J+1/2)^2/2\sigma^2},
\end{equation}
with the spin-cutoff parameter $\sigma$ was assumed. We focused solely
on the description of the spin distribution using an empirical,
energy-independent expression for the spin cutoff for a nucleus with mass
$A$~\cite{Gar01,Ign93}:
\begin{equation}
\sigma=(0.98 \pm 0.23)A^{(0.29 \pm 0.06)},
\end{equation}
yielding $\sigma=3.08$ for $A=52$. Fig.~\ref{fig:density} presents the
spin frequency from the shell-model calculations using the GXPF1
effective interaction in comparison to the statistical approach. A
good agreement is reached with this parameter-free description.

In summary, the odd-odd $fp$-shell nucleus \nuc{52}{Sc} was
investigated with in-beam $\gamma$-ray spectroscopy following
fragmentation of \nuc{55}{V} and \nuc{57}{Cr}. A new $\gamma$-ray
transition was observed at 212(3)~keV and was assigned to the
low-lying level scheme of \nuc{52}{Sc}. All known states were compared
to full $fp$-shell calculations with the GXPF1 effective interaction.
The placement of the new $\gamma$ decay was also discussed in
comparison to shell-model calculations with the GXPF1A and KB3G
effective interactions. All three interactions predict a very similar
low-lying level scheme, illustrating the predictive power of these
more modern interactions for the $fp$ shell. The frequency of spin
values from the GXPF1 shell-model calculation was successfully modeled
by the spin distribution formulated in a purely statistical approach
using an empirical, energy-independent spin-cutoff parameter that is
only a function of the mass number $A$.

We thank T.\ Baumann, T.\ Ginter, M.\ Portillo, A.\ Stolz, and the
NSCL cyclotron operations group for the high-quality secondary and
primary beams. This work was supported by the National Science
Foundation under Grants No. PHY-0110253 and PHY-0244453, and by the
U.S. Department of Energy, Nuclear Physics Division, under Contract
Nos. W31-109-ENG-38.

\end{document}